\documentclass[conference]{IEEEtran}
\IEEEoverridecommandlockouts
\usepackage{cite}
\usepackage{amsmath,amssymb,amsfonts}
\usepackage{algorithmic}
\usepackage{graphicx}
\usepackage{xcolor}
\usepackage{textcomp}
\usepackage{caption}
\usepackage{subcaption}
\usepackage{soul}
\usepackage{flushend}
\usepackage{comment}
\usepackage[english]{babel}
\usepackage[utf8]{inputenc}
\usepackage{fancyhdr}
\usepackage{textpos}
\usepackage{eso-pic}

\begin{document}	
	
\AddToShipoutPictureBG*{%
	
	\AtPageUpperLeft{%
		\setlength\unitlength{1in}
		\hspace*{\dimexpr0.5\paperwidth\relax}
		\makebox(0,- 0.5)[c]{\small This is the authors’version of the paper that has been accepted for publication in IEEE ICC 2018}%
}}

\AddToShipoutPictureBG*{%
	
	\AtPageUpperLeft{%
		\setlength\unitlength{2in}
		\hspace*{\dimexpr0.5\paperwidth\relax}
		\makebox(0,- 0.5)[c]{\small Communication and Information Systems Security Symposium, 20-24 May 2018, Kansas City, USA}%
}}

\title{Cross-Layer Authentication Protocol Design for Ultra-Dense 5G  HetNets}

\author{\IEEEauthorblockN{Christian Miranda Moreira}
	\IEEEauthorblockA{\textit{Electrical Engineering Department}\\
		\textit{\'Ecole de Technologie Sup\'erieur}e\\
		Montr\'eal, Canada \\
		christian.miranda-moreira.1@ens.etsmtl.ca}
	\and
	\IEEEauthorblockN{Georges Kaddoum}
	\IEEEauthorblockA{\textit{Electrical Engineering Department}\\
		\textit{\'Ecole de Technologie Sup\'erieure}\\
		Montr\'eal, Canada \\
		georges.kaddoum@etsmtl.ca}
	\and	
	
	\IEEEauthorblockN{Elias Bou-Harb}
	\IEEEauthorblockA{\textit{Cyber Threat Intelligence Laboratory}\\
		\textit{Florida Atlantic University}\\
		Florida, USA \\
		ebouharb@fau.edu}}
\maketitle
\begin{abstract}
Creating a secure environment for communications is becoming a significantly challenging task in 5G Heterogeneous Networks (HetNets) given the stringent latency and high capacity requirements of 5G networks. This is particularly factual knowing that the infrastructure tends to be highly diversified especially with the continuous deployment of small cells. In fact, frequent handovers in these cells introduce unnecessarily recurring authentications leading to increased latency. \\In this paper, we propose a software-defined wireless network (SDWN)-enabled fast cross-authentication scheme which combines non-cryptographic and cryptographic algorithms to address the challenges of latency and weak security. Initially, the received radio signal strength vectors at the mobile terminal (MT) is used as a fingerprinting source to generate an unpredictable secret key. Subsequently, a cryptographic mechanism based upon the authentication and key agreement protocol by employing the generated secret key is performed in order to improve the confidentiality and integrity of the authentication handover. Further, we propose a radio trusted zone database aiming to enhance the frequent authentication of radio devices which are present in the network. In order to reduce recurring authentications, a given covered area is divided into trusted zones where each zone contains more than one small cell, thus permitting the MT to initiate a single authentication request per zone, even if it keeps roaming between different cells. Accordingly, once the RSS vectors and the encrypted mobile identification are received by the authentication slice (AS), this latter builds the authentication vector using the $k$-nearest neighborhood technique to estimate the $k^{\rm dh}$ fingerprint distribution which is compared to the radio trusted zones database to prove the legitimacy of the MT and the network slice (NS). Cross-layer authentication protocol is consequently executed. The proposed scheme is analyzed under different attack scenarios and its complexity is compared with cryptographic and non-cryptographic approaches to demonstrate its security resilience and computational efficiency. \\

\end{abstract}

\begin{IEEEkeywords}
Cross-layer authentication protocol, Recommendation system, Radio trusted zone database, $k$-NN, SDWN.
\end{IEEEkeywords}

\section{Introduction}

Wireless connectivity has progressively secured its place in the last decade to be an indispensable part of our communication means that has undoubtedly increased  mobile traffic load. According to \cite{19}, mobile data traffic is expected to expand at a compound annual rate of 57\% until 2019 and is predicted, by year 2020, to exhaust the available capacities provided by the fourth generation (4G) and the long term evolution (LTE) infrastructures \cite{2}.

In addition, network densification using low-power small cells is considered to be a core solution for 5G. 
This new architecture indeed demands new requirements such as flexibility in management and configuration, adaptability and vendor-independence. To meet these requirements, software defined wireless networks (SDWN) have been proposed as a cost-effective solution \cite{20}.
Hence, the hetnet nature of 5G with the separation of data and control planes and the virtualization of major network functions increase the need for authentication improvement, integrity, and privacy protection in the presence of malicious actors \cite{22}.

The traditional authentication handover mechanism is based on a cryptographic key and on multiple handshakes. 
The authentication and key agreement (AKA) protocol which is standardized by the third generation partnership project (3GPP) in \cite{6} is widely used in current wireless networks. In brief, the AKA protocol involves three entities which are ({\it i}) the mobile terminal (MT) which represents the user, ({\it ii}) the home environment (HE) and ({\it iii}) the serving network (SN).  

AKA allows the SN to authenticate and exchange keys with the user, without ever being given the user's key. Instead, one-time authentication vector (AV) are issued to SN by the HE. All communication and computations in AKA are very efficient thanks to the use of symmetric-key cryptography.  To this end, the client authenticates the network by computing the response (RES) using its k secret key and the network authenticates back to the client across-AV by associating its response with the expected response (XRES). Using symmetric cryptography, AKA shares a k secret key with the MT and the HE in order to maintain the privacy and security of the information.\\


By exploring and investigating the security analysis of current authentication protocols, we pinpoint several of their vulnerabilities against different attacks including resistance attacks, black hole attacks, replay attacks, man-in-the-middle attacks, impersonation attacks, and denial of service attacks \cite{22}. We also note that considerable research has been made to improve such protocol. For instance, in \cite{5}, the authors propose a security enhanced authentication and key agreement (SE-EPS AKA) method based on wireless public key infrastructure by using the ellipse curve cipher (ECC) encryption. Additionally, the research work in \cite{4} points out a scheme which resolves the privacy problem and prevents mobility management entity (MME) masquerading. Moreover, the devised scheme takes into consideration the fact that the MT is energy-limited and for that reason, public key cryptography is not used at the MT. The mechanism in \cite{7} suggests an enhanced AKA protocol using a methodology which provides zero-knowledge proof using a pre-shared key that is never sent over the transmission medium. A new key exchange procedure is proposed in \cite{6} where the user identity information and authentication vector in the network domain are encrypted using the public key cryptosystem. The public parent key adopted in local authentication is generated by means of random data. In \cite{8}, the authors show that their proposed approach eliminates the synchronization between mobile station and its home network in the key exchange process. Besides the discussed vulnerabilities, the conventional AKA authentication protocol may not fulfill the requirements of future dense small 5G network cells in terms of security, resistance to spoofing, low latency, infrequent handover and low computational costs \cite{10}.

Alternatively, it has been shown that exploiting the environment-dependent radiometric features of a specific transceiver pair, such as the channel state information (CSI) \cite{15} and the received signal strength indicator (RSS) \cite{18}, can improve the authentication procedure. In fact, these channel characteristics can be used to differentiate signals arriving from authorized transmitters and those originating from spoofed transmitters \cite{11}, \cite{12}. Moreover, \cite{16} presents a comparative survey of wireless local area network location fingerprinting schemes. The foundation behind these schemes is that RSS is location-specific, due to path loss and channel fading, where most works in this category usually assume that the users are static; thus generating an excessive false positive rate in mobile scenarios. Accordingly, an attacker who is at a different location from the genuine user might be placed in different RSS profiles and whereby can infer the RSS of the user by using a wireless sniffer tool. 

To tackle these challenges, a promising cross-layer authentication method is proposed in this paper. The novelty of our work lies in devising and evaluating a multi-layer approach which amalgamates physical layer information (i.e., non-cryptographic) \cite{14} in conjunction with cryptographic procedures. In this context, we define two security level agreements (SLAs) which are devised for decentralized and centralized networks, respectively.  These agreements are established at the beginning between the network slice (NS) and the authentication slice (AS). \\

Moreover, we prosed the use of   a radio trusted zones data base at the AS side. In fact, a  given covered area is divided into trusted zones where each zone contains more than one small cell, thus permitting the MT to initiate a single authentication request per zone, even if it keeps roaming between different cells. On the other hand, the data base  of each zone  contains the different RSS profiles and their corresponding  localizations. Thanks to the widely used radio mapping technique, this database is filled. Hence, this approach aims to add another security level to the system and reduce the recurring authentications in the network.

At the MI side and for non-cryptographic procedures, the gathered RSS measurements at the MT are used to generate the  $k^{\rm th}$ fingerprint aiming to randomize the secret key used by the AKA protocol. After this step, a cryptographic approach employing an enhanced AKA protocol is performed in order to improve the confidentiality and integrity of the authentication handover. Furthermore, sending the mobile identity (IM) on the fly in a clear form (without encryption) is still another weaknesses of the AKA protocol. We address this problem by generating a radio signal fingerprint that prevents such transmission patterns, thus obscuring IM. Subsequently, the obscured IM is encrypted and then transmitted with the RSS parameter to the AS to corroborate the MT identity within the NS.

Once received, the AS, in response, sends an AV built with the aid of the $k^{\rm th}$ fingerprint, to approve the NS identity into the MT. In addition, before AS sends AV to MT across NS, AS applies the $k$-nearest neighborhood ($k$-NN) technique on the revived RSS with the existing data base to estimate the $k^{\rm dh}$ fingerprint distribution. The nearest output of this algorithm is used to identify the corresponding legitimate location stored in the database.
   It should be mentioned that the inaccuracy of the estimation technique does not affect the reliability of the proposed protocol because the identification of legitimate location is  already takes into consideration the localization error range to define the trusted radio zones. 
 Finally, the AKA authentication protocol is performed as operated in conventional cryptographic protocols.\\

To have better insights into this work, we frame the set of contributions of this paper as follows:

\begin{enumerate}
	
	\item Defining two SLA for decentralized and centralized 5G networks and proposing a cross-layer authentication approach based on SLA specifications.
	
	\item Exploiting the random and unique RSS \emph{measurements} in order to compute a secret $k^{\rm th}$ fingerprint.
	
		\item Enhancing the security level of 5G networks by introducing a novel approach rendered by the creation of a radio trusted Zones database.
		
	\item Executing security protocol analysis and validating the sensitivity of the proposed cross-layer protocol against different threats by leveraging the AVISPA tool. Additionally, comparisons of the computational complexity of the proposed scheme against traditional cryptographic and non-cryptographic approaches are also conducted.\\

\end{enumerate} 
To  the  best  of  author's knowledge, the cross-authentication approach along with radio trusted zones have not yet been devised and evaluated in the literature. The remainder of this paper is organized as follows.  Section II  introduces the proposed system model and the protocol design. In Section III, the security and performance analysis are evaluated. Finally, this paper is summarized in Section IV, where a number of future endeavors are also put forward. 

\section{System Model and Protocol Design}
In order to tackle the important security challenge in SDWN-based 5G HetNets which results from the separation of the radio control plane from the data plane, we propose an AS as a third party security agent to provide isolation and efficient security authentication management over the integral network. Therefore, a cross-layer authentication procedure is proposed. This procedure is mainly based on increasing the security level of the AKA protocol by using physical layer information and machine learning algorithms at the server side in order to estimate the authenticity of the radio devices. The following subsections will detail the various steps related to the proposed protocol, namely, fingerprint generation, estimation and distribution, the cross-layer authentication and protocol design.


\subsection{Generation of the $k^{\rm th}$ fingerprint} 
\begin{figure}[h]
	\centering
	\includegraphics[width=0.5\textwidth]{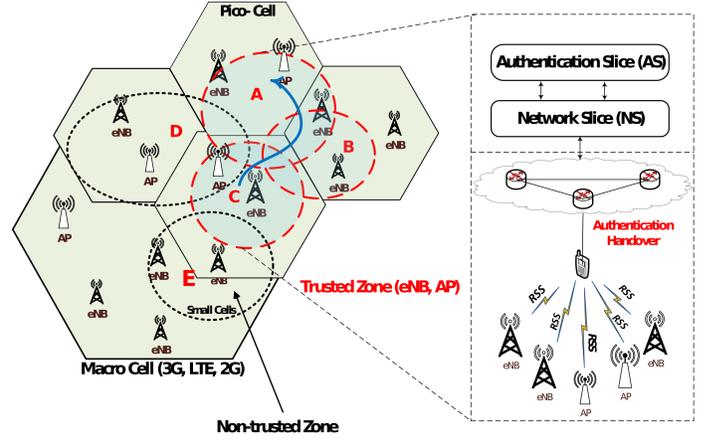}
	\caption{RSS vectors transfer
		between 5G radio devices through the MT in a SDWN Architecture}\label{fig:fig2}
\end{figure}

In our approach, we employ a channel-based fingerprinting mechanism to enhance the authentication procedure. Towards this end, we first define two SLAs which address decentralized and centralized networks, respectively. For decentralized networks, the authentication procedure is comprised of two steps. For centralized networks, a complete three steps approach is applied. These agreements are established at the beginning between the NS and the AS. \\ After defining the agreement, the non-cryptographic procedure is performed. As shown in Figure \ref{fig:fig2}, the RSS measurements from different base stations are gathered and then averaged. In fact, to make this RSS parameter unique and random to suit the generation of the $k^{\rm th}$ fingerprint key, different RSS values from various radio devices are required to compute the average. Otherwise, considering a single RSS measurement from one radio device and due to the multi-path propagation environment, two different users on different locations may have the same value, which hinders the security of the protocol.  Hence, the received radio signal strength from different radio devices, when collected at the $u^{th}$  MT side, can be represented as
\begin{equation}
{\bf RSS}_{{\bf}u}  = [R_{1,t_1},R_{2,t_2} \ldots, R_{N,t_n}], 
\end{equation}
where $t_i$ is the \emph{time of arrival} of the signal received from the $i^{th}$ access point $R_{i,t_i}$ to the $u^{th}$ MT at a given location. This time of arrival significantly reduces the possibility to impersonate the RSS vectors by an intruder.

The MT then averages the RSS vectors to generate the $k^{\rm th}$ fingerprint such that
\begin{equation}
\label{equationav}
k =  \rm{E}{[{\bf RSS}_{{\bf}u}  }],
\end{equation}
where $\rm{E}[.]$ is the mean operator. 

The generated $k^{\rm th}$ fingerprint aids in randomizing the secret key that is used by the AKA protocol.
After this step,  the AKA protocol is performed at the MT.
As a first step in this protocol, the IM is masqueraded by the $k^{\rm th}$ fingerprint. The output of the masquerading, dubbed as temporary identification mobile (TIM) aims to hide the device IM.
After masquerading, in order to protect TIM from catching attack, 
this latter is encrypted with the AES ecryption algorithm. 
Finally, MT sends TIM with the RSS vectors to the AS to corroborate the MT identity within the NS.
\subsection{Estimation of the $k^{\rm dh}$ fingerprint distribution} 
In this section, we will first introduce the proposed radio trusted zone concept that we consider in our system design to recognize the legitimacy of different radio device identities in the network. To this end, each zone is set to form a cluster of neighboring small cells. The database of this latter is built thanks to radio map database using the localization fingerprinting  method in \cite{16}.

Since building the radio trusted zone database is out of the scope of this paper, in the remaining of this work, we assume the existence of this database at the AS side. Once the radio signal is received (i.e., TIM and RSS vectors), the AS analysis the RSS vectors and computes the $k^{\rm th}$ fingerprint as given in Eq. (\ref{equationav}). The resultant key is used to  unmask TIM in order to corroborate the IM authenticity within the NS. After this step, the deterministic $k$-NN method  is used to estimate the $k^{\rm dh}$ fingerprint distribution. In fact, the $k$-NN method is one of the simplest ways to determine the fingerprinting process of wireless devices by using a radio map database. Hence, in contrary to our solution, the conventional $k$-NN method is victim of  false positive alarms when its output is compared to a  radio trusted zone  without taking into account the localization error range. Finally, in the proposed system, the $k$-NN process considers multiple nearest neighbors to compute the $k^{\rm dh}$ fingerprint distribution as follows
\begin{equation} \label{knn}
 k^{dh}=  \min \sqrt {\sum\limits_{y = 1}^Y {\left( {{\bf RSS}_u  - {\bf RSS}_y } \right)^2 } },
\end{equation}
\\
where $ {{\bf RSS}_y }$ is the RSS vector stored in the radio trusted zone. The resultant $k^{\rm dh}$ is used to identify the corresponding location in the trusted radio zones database, taking into account the  localization error range, to prove the legitimacy of the radio device. In the next subsection, we will detail the key exchange process; handover through the cross-layer authentication protocol. 





\subsection{Cross-layer Authentication Protocol}



\begin{figure}[!t]	
	\centering 
	\includegraphics[width=0.5\textwidth]{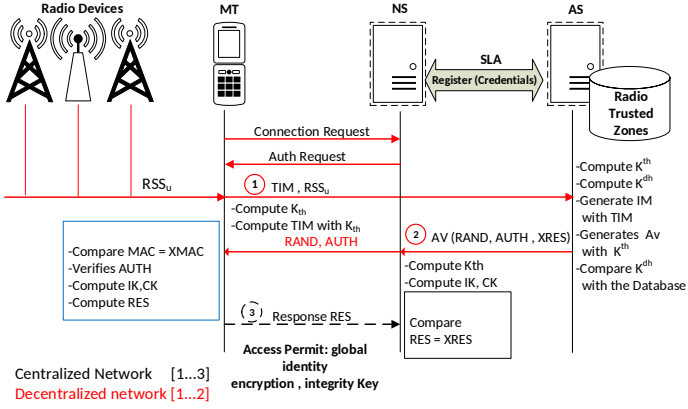}
	\caption{Cross-layer authentication handover procedure}\label{fig:fig3}
\end{figure} 

The procedure of keys' exchange between different network entities is exhibited in Figure \ref{fig:fig3}. In this context, we assume that NS possesses previous credentials to coordinate with the AS. The following steps within the authentication protocol are executed only if the MT is a new user or enters a certain trusted zone for the first time.
%
%



\begin{description}  
	\item{Step 1:}
	\begin{itemize}	
	
	\item{MT computes the $k^{\rm th}$ fingerprint based on Eq. \eqref{equationav} and      
		masquerades IM with the $k^{\rm th}$ resulting TIM.}
	\item{MT sends TIM and RSS vectors to AS in response to the demand made by the NS.} 
	\end{itemize}

	\item{Step 2:}
	\begin{itemize}	
	\item{AS generates the $k^{\rm th}$ fingerprint based on RSS vectors and IM from TIM, and $k^{\rm th}$}.
	\item{AS estimates $k^{\rm dh}$ fingerprint distribution and search its corresponding legitimate localization in the radio trusted zones database that has been previously built.}
	\item{AS generates AV only if the legitimate localisation is found in the database}.
    \end{itemize}

\end{description}

Hence, AV will contain the following keys

\begin{center}	
	
	\begin{itemize}			
		\item ${\rm MAC}$ = $f_1$( $k^{\rm th}$, ${\rm AMF}$, ${\rm SQN}$, ${\rm RAND}$)  \\
		\item ${\rm XRES}$ = $f_2$( $k^{\rm th}$, ${\rm RAND}$) \\
		\item ${\rm CK}$ = $f_3$( $k^{\rm th}$, ${\rm RAND}$)\\
		\item ${\rm IK}$ = $f_4$( $k^{\rm th}$, ${\rm RAND}$)\\
		\item ${\rm AK}$ = $f_5$( $k^{\rm th}$, ${\rm RAND}$)  \\
		\item ${\rm AUTH}$ = ${\rm SQN} \oplus {\rm AK}$$||$${\rm AMF}$$||$${\rm MAC}$\\
	\end{itemize}
\end{center}
where  $ \oplus$ and $||$ denote the bitwise XOR and the concatenation operations, respectively. The notions $f_1$ to $f_5$ are the AES cryptographic hash functions, SQN is the fresh sequence number, AMF denotes a public authentication management field handled by the network operator, RAND signifies a random number, IK symbolizes the integrity key, CK refers to the cipher key, AK is the anonymity key, MAC denotes message authentication code, XRES is the expected response, XMAC is the expected MAC and AUTN implies the authentication token. \\

It is important to note  that for the decentralized network, the cross layer authentication algorithm ceases in step 2 thus AS sends the AV[RAND, AUTH] to MT without passing by NS, then MT verifies SQN and compares XMAC with the  MAC to validate the network. Therefore, for the case of the centralized network, the AS sends AV[RAND, AUTH, XRES] to the NS and then the NS sends AV[RAND, AUTH] to MT. Subsequently, the MT calculates different keys as follows:

\begin{flushleft}
	\begin{itemize}  	
		\item ${\rm AK}$ = $f_5$( $k^{\rm th}$,  ${\rm RAND}$) \\
		\item ${\rm SQN}$= 1st(${\rm AUTN}$) $\oplus$ ${\rm AK}$ \\ 
		\item ${\rm XMAC}$= $f_1$($k^{\rm th}$, 2nd(${\rm AUTN}$), ${\rm SQN}$, ${\rm RAND}$)\\
		\item ${\rm RES}$ = $f_2$( $k^{\rm th}$, ${\rm RAND}$) \\
		\item ${\rm CK}$ = $f_3$( $k^{\rm th}$, ${\rm RAND}$)  \\
		\item ${\rm IK}$ = $f_4$( $k^{\rm th}$, ${\rm RAND}$)\\
	\end{itemize}
\end{flushleft}

After calculating the keys, the following is performed:

\begin{description}  
	
	\item{Step 3:}
	\begin{itemize}	
	\item{ If SQN is in the correct range, then the XMAC is compared with MAC to validate the network.}
	\item{ If SQN is not in the correct range, then the connection is rejected.}
	\item{ Once validated, MT calculates its RES value and sends it to the NS for validation.}
	\item{NS compares RES with XRES that is already present in the authentication vector.}
	\item{If RES is equal to XRES, then the MT is also authenticated by NS and mutual authentication is achieved. Otherwise, it is rejected.}
	
	\end{itemize}
\end{description}  	

\section{Security and Performance Analyses}

In the following, we perform automated security analysis to asses the security level of the proposed cross-layer authentication protocol using the Avispa tool to verify its resistance against various attacks. Moreover, we leverage a JAVA API to estimate the computational cost of the proposed scheme. 

\subsection{Security Analysis} 
 In this section, we analyze the security of the conventional authentication protocol and the proposed authentication protocol. In the first scenario, we assess the security of the conventional AKA protocol. Since this protocol sends its pre-shared key over the air, we consider herein that an intruder in the network has the knowledge about the key. In contrast, in the second scenario related to the proposed protocol, the intruder is unable to acquire knowledge about the pre-shared key given that this protocol does not send the key over the air. In both scenarios, the intruder performs several typical attacks (i.e., man-in-the-middle, redirection, replay, etc.) on the protocols. \\
 
\subsubsection{Scenario 1}
\begin{figure}[h]		
	\centering
	\includegraphics[width=0.4\textwidth]{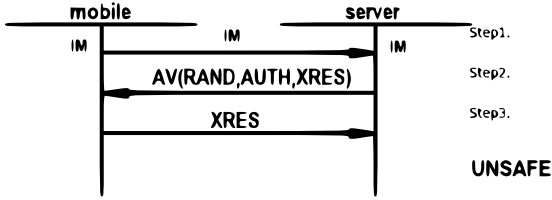}
	\caption{Avispa simulation for scenario 1}\label{fig:fig8}
\end{figure}

In this scenario, the MT sends the IM and the secret key k on the fly to AS to initiate the authentication process. This process is formalized and then assessed using Avispa tool, As depicted in Figure  \ref{fig:fig8} the protocol analysis indicates \textbf{UNSAFE}, revealing that the protocol is vulnerable to various analyzed threats.
  
\subsubsection{ Scenario 2} 
\begin{figure}[h]	
	\centering
	\includegraphics[width=0.4\textwidth]{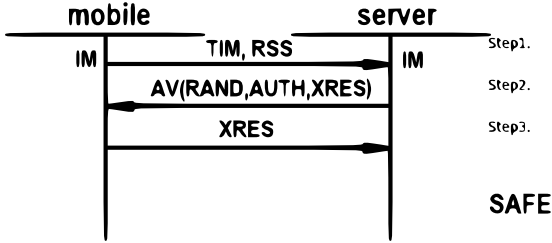}
	\caption{Avispa simulation for scenario 2} \label{fig:fig10}
\end{figure}

As described in our protocol, MT sends IM encrypted with  $RSS_u$  on the fly to AS. In contrast to the conventional mechanism, MT, SN and AS generate the $k^{\rm th}$ fingerprint separately which improve the security as the fingerprint is never sent on the fly. This is corroborated by conducting protocol analysis using Avispa tool, which indicates that this protocol is \textbf{SAFE} (against the analyzed threats) as shown in Figure \ref{fig:fig10}.


In the following, we detail how different attacks could be performed under scenario 2 and how our protocol design is resilient against such threats.\\

\paragraph{Redirection attacks and black-hole attacks}
The mobile identification is not protected in the current mobile network and can be altered by an adversary with some devices such as an IM catcher, which leads to the redirection attack. In our protocol, the $k^{\rm th}$ fingerprint is used to masquerade the IM and thereby protects 5G networks against redirection attacks. 

Accordingly, the attack fails if the malicious user is unable to obtain the legitimate user information from the MT. In the proposed protocol, the MT computes the IM embedded with the $k^{\rm th}$  fingerprint  generating TIM and sends it to the AS. The authentication request is denied if the AS fails to match the IM sent by the MT. Such a technique solves the problem of miss-charged billing in the 5G network. Thus, the proposed scheme immunes the 5G network from black-hole attacks.\\

\paragraph{Replay attacks}
The cross-layer protocol is resilient against this attack by solely sending the RSS vectors and TIM during the transmission of information over the network. This prevents the misuse of valid information; an adversary typically can delay the message over the network and sends it later for some malicious purpose if no random number or fingerprint is involved in the transmitted message.\\

\paragraph {Man-in-the-middle attacks}
A man-in-the-middle attack occurs when an adversary eavesdrops the communicated information between the MT and the NS. In the context of the our proposed cross-layer protocol, the $k^{\rm th}$ key is independently generated in the MT, AS and NS. This key prohibits the communication from being eavesdropped. \\

\paragraph{Impersonation attacks} 
Over the 5G network, the corruption of the control plane endangers the security of the whole network. Following are some scenarios in which an adversary may attempt to impersonate the 5G network.\\

\begin{description}  
	\item[Case 1:] \hspace{5pt} Consider the presence of a fake NS where an intruder can eavesdrop all its messages. The adversary must reply with a valid response RES to the NS in order to impersonate the MT, but the intruder cannot obtain the correct RES since this latter is exchanged exclusively between the MT and an uncorrupted AS. 
	
	\item[Case 2:] \hspace{5pt} If the intruder attempts to impersonate an uncorrupted network, the attempt would fail as the MT can verify that previously, there was no initiated request for AV. Furthermore,  MT only exchanges traffic with trusted radio devices (i.e., the radio trusted zones database). \\
	
\end{description}  

\paragraph{Denial of Service (DoS) attacks}

The DoS attack and its variants are discussed in the following scenarios; the attacker MT's flood the victim control plane with authentication requests by spoofing the IM/TIM, the k key and a request number.

\begin{description}
	\item[Case 1:] \hspace{5pt} The attacker MT floods the NS victim with self IM. If the malicious MT does not respond within the threshold time duration to the proxy, then the connection is simply terminated. Accordingly, NS resets the authentication request and releases the resources that are used to maintain the authentication request status. In addition, if the request is originating from a malicious user, then the proxy will not acquire the $k^{\rm th}$ key or would simply receive an invalid $k^{\rm th}$ key. There is a timeout period for each MT to maintain the state of half-opened authentication requests. If the malicious MT attempts to cause an overflow at the victim NS with the half-open authentication requests, NS would not be able to accept any new incoming authentication requests.	

	\item[Case 2:]  \hspace{5pt} The attacker MT floods the victim NS by spoofing IM. In this scenario, if the actual MT that receives a message is not active, then the AS will not receive any information from the MT, and this process becomes similar to first case; the NS waits for a threshold time to hear from the AS. After the timeout period, the NS resets the authentication request and releases the resources that are used to maintain the authentication request status.

In fact, in this protocol, the AS is supposed to receive an RSS vectors from the MT, which is neither an actual IM of the MT nor a TIM for the NS. An actual IM or  TIM with a fake $k^{\rm th}$ key will not be able to extract the correct IM of the MT and thus the connection will be terminated. Hence, there is no chance that the attacker would be able to generate the same $k^{\rm th}$ from a victim MT's IM. Indeed, given the aforementioned information, we assert that the proposed cross-layer AKA protocol protects the network from DoS attacks.
	
\end{description}

\subsection {Computational cost analysis}
We further thought that it would be insightful to analyze the computational cost of our proposed cross-layer protocol.
In this context, it is important to note that the SDWN paradigm introduces the cloud radio access networks (C-RAN) paradigm,  which  aims at reducing the computational cost as most of the processing activities are executed on the distributed cloud. Moreover, the well-trusted radio zones database is formed by small cells; a mechanism which avoids frequent authentication of the MT within each small cell.

We perform comparisons of the non-cryptographic and cryptographic authentication algorithms against the proposed cross-layer protocol 
For our analysis, we exploit a dataset of 25 radio signal strength samples collected for 280 combinations of user locations and orientations. 

Since our proposed cross-layer protocol is implemented in Java, a Java API is developed for this evaluation purpose to be coherent and to generate real time perspectives of the computational cost. Moreover, we use an Intel Core i7-6700 CPU with 3.4 GHz X64 based processor and 16 GB RAM to conduct the computations. 

\begin{figure}[!t]	
	\centering
	\includegraphics[width=0.5\textwidth]{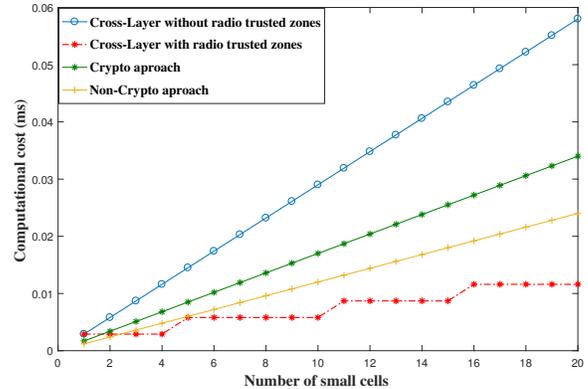}
	\caption{Cross-layer authentication protocol with and without radio trusted zones, in comparisons with cryptographic and non-cryptographic approaches}
	\label{fig:fig5}
\end{figure}

The results of this comparison is shown in Figure \ref{fig:fig5}, which demonstrates the computational cost of cryptographic, non-cryptographic and the proposed cross-layer protocol across small cells with and  without employing the trusted zone approach at the AS level. In the case where the proposed protocol operates without employing the radio trusted zones, we observe a clear increment in computational cost in comparison with non-cryptographic and cryptographic procedures, respectively, and this gap increases when the number of cells increases. The augmentation is due to the fact that our proposed cross-layer protocol operates in this specific case without a trusted zone and uses machine learning algorithm to authenticate the radio devices. The absence of a trusted zone increases the recurrence of authentication procedures, which leads to more complexity.

%

Once the radio trusted zones' approach is employed, the computational cost of the proposed cross-layer protocol drops in contrast with the first approach. This renders the deployment of a radio trusted zones a better choice to achieve a lower complexity and thus reduce latency. \\


\section{Conclusion}
In this paper, we propose a software-defined wireless network (SDWN)-enabled fast cross authentication scheme that combines non-cryptographic and cryptographic algorithms to tackle the challenges of latency and weak security in 5G HetNets.  First, the radio trusted zone database concept is introduced aiming to reduce the authentication recurrence.  Consequently, the cross-layer algorithm is designed, implemented and evaluated. By executing automated protocol analysis using the Avispa environment, the security posture of our cross-layer authentication protocol in terms of resilience to various attacks is analyzed. The results show that the  proposed  scheme satisfies 5G security requirements and its advantages have been verified by simulations. Further, the  proposed  protocol  causes considerable deduction of traffic authentications, thanks to the introduction of the radio trusted zone unit. Finally, a Java API is developed to compute the complexity of our system and to compare it against cryptographic and non-cryptographic approaches. It is shown that if a radio trusted zone is employed, the computation complexity is significantly reduced in comparisons with the two latter approaches, by limiting the authentication recurrence. As for future work, we will be focusing on employing  machine learning techniques to properly classify the various RSS profiles of a HetNet in an attempt to build reliable and efficient radio trusted zones.

\bibliographystyle{IEEEtran}
\bibliography{ICC_2018}
 
\end{document}